# 10 Inventions on improving keyboard efficiency
## -A TRIZ based analysis

**Umakant Mishra**
Bangalore, India
umakant@trizsite.tk
http://umakant.trizsite.tk

**Contents**

# 1. Introduction

A keyboard is the most important input device for a computer. With the development of technology a basic keyboard cannot remain confined within the basic functionalities of a keyboard, rather it has to go beyond. There are several inventions which attempt to improve the efficiency of a conventional keyboard.

## 2. Inventions on keyboard efficiency

### 2.1 Computer keyboard (Patent 5600313)

**Background problem**

A keyboard is used as an input device for the computer. Although some keyboards contain function keys, their functions are different from software to software and hence difficult for the user to remember. There are no special keys on the keyboard to be used for specific jobs like printing, saving and other common tasks. There is a need for a keyboard, which contains predefined keys for various common functions.

**Solution provided by the invention**

Lorri Fredman invented (patent 5600313, issued Feb 1997) a keyboard that provides several mouse buttons on the keyboard. A set of static icon keys is positioned vertically on the left side of the keyboard. The command icon keys are positioned vertically on a right side of the keyboard. The tool bar icons are positioned horizontally above the function keys of the keyboard. These three sets of icon keys free up display screen which allows more screen space for the application.

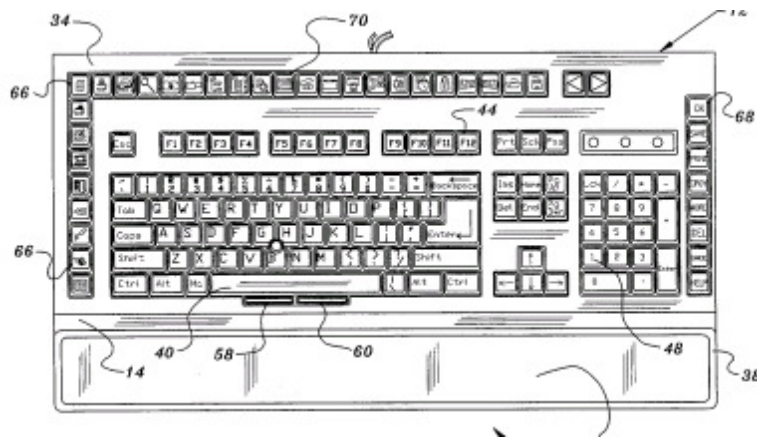

This keyboard provides all the advantages of a standard keyboard, besides it also provides the additional advantages of special icon keys.

**TRIZ based analysis**

The invention substitutes the visual toolbar icons with tactile keys on the keyboard **(Principle-28: Mechanics substitution).**

The new keyboard provides additional keys over and above all the conventional keys **(Principle-38:Improve quality)**.

## 2.2 Power saving method for Computer keyboard (Patent 5767594)

**Background problem**

A conventional computer keyboard consumes electric power continuously as long as the computer is running irrespective of its actual usage. Even if there is no action on the keyboard, still it consumes electricity. This is a waste of energy. The method needs to be improved.

**Solution provided by the invention**

Andrew Cheng disclosed the same mechanism (Patent 5767594, assigned to Holtek Microelectronics, Issued in June 98) to save power consumption for the computer keyboard.

As per the invention, the keyboard power saving method includes sensing if there is keystroke action taking place. When time period of no keystroke action exceeds a pre-set threshold, power supply to the keyboard will be shut down automatically. When a keystroke action is detected, power supply to the keyboard will be resumed instantly. Thus can save energy when keyboard is idle from time to time.

**TRIZ based analysis**

The keyboard should consume no power **(Ideal Final Result)**.

The keyboard should not consume power when not used **(desired result)**.

The invention monitors the keystroke action on the keyboard and shutdown keyboard when there is no action on the keyboard **(Principle-23: Feedback).**

## 2.3 Computer keyboard with switchable typing/ cursor control modes (Patent 5864334)

**Background problem**

There are different cursor control mechanisms like using a mouse, a trackball, a touchpad and so on. All of them have their own advantages and disadvantages.
The mouse requires significant amount of desktop space. The trackball is integrated into the keyboard area in a fixed location and is not convenient to many users. A touchpad has a relatively small surface area and inconvenient for long cursor movements. A pointing stick is very small for many users to locate and operate.

**Solution provided by the invention**

Sellers invented a method of cursor control (patent 5864334, Assigned to Compaq, Issued in Jan 99) by using an optical mechanism. The invention discloses a toggle switch on the keyboard which will change the keyboard mode to cursor control mode. The cursor control mode will activate the video camera mounted above the keyboard to monitor the movement of user's hand on the keyboard. When the user moves his finger on the keyboard, the cursor is moved accordingly on the computer screen. This video controlled cursor positioning system can be used in laptops and desktop computers.

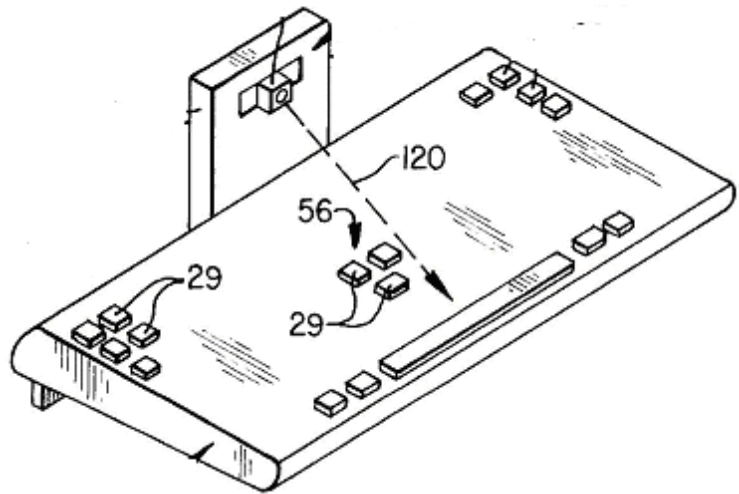

**TRIZ based analysis**

There should be no requirement of an extra pointing device. Keyboard should work as a full featured pointing device **(Ideal Final Result)**.

The invention uses use the same keyboard for keyboard input and pointing device both the purposes **(Principle-6: Universality)**.

The invention uses an optical mechanism to detect the finger movements for cursor control **(Principle-28: Mechanics Substitution)**.

**2.4 Computer bottom keyboard incorporating arrangement for enhanced cooling (Patent 5978215)**

**Background problem**

The portable computers are compact in size and, unlike desktop PCs, have no space for ventilation. This makes the notebook get heated quickly. One of the methods to reduce heat is using high power cooling fans. But unfortunately they require more battery consumption. There is need to increase the cooling capacity of the portable computers.

**Solution provided by the invention**

Chiu et al. disclosed a keyboard for portable computers having enhanced cooling mechanism (patent 5978215, assigned to IBM, Nov 99). According to the invention the portable computer has outwardly slidable compartments above the keyboard. When these compartments are slid out, the structure provides increased surface areas for dissipation of heat.

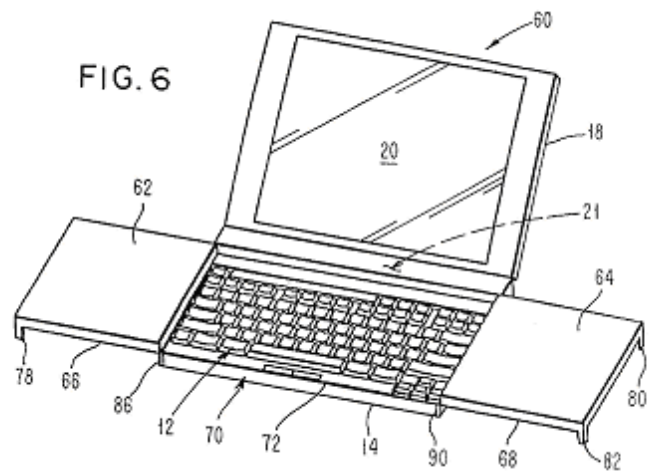

### TRIZ based analysis

The portable computer should automatically cool by itself **(Ideal Final Result)**.

The invention exposes more internal components of the computer for faster dissipation of heat **(Principle-35: Parameter change)**.

One of methods used by this invention is an openable or slidable laptop body **(Principle-15: Dynamize)**.

### 2.5 Air ventilation holes on a keyboard (Patent 5982615)

### Background problem
The CPU of the computer generates a lot of heat during operation which can go even more than 100 degree centigrade. This temperature is harmful for the computer and might disable computer operation. Moreover laptops, unlike desktop computers, do not have enough space for ventilation. It's necessary to cool down the CPU of a portable computer. How to control the heat in laptops?

### Solution provided by the invention
Kwang-Ho Song invented a ventilating keyboard (US patent 5982615, assigned to Samsung Electronics, Nov 99). According to the invention the keyboard will have air ventilation holes. The heat-emitting device will ventilate the heat to outside though the air ventilation holes on the keyboard. This technique increases the cooling effect of a portable computer.

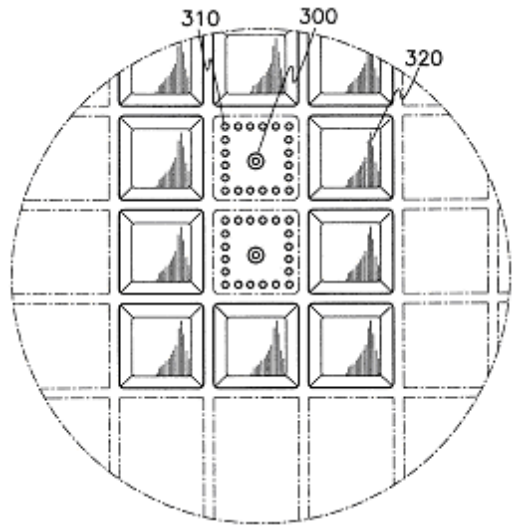

Holes through keyboard does not allow dust or external particles while carrying the laptop, as the keyboard is normally covered with the display when carried.

**TRIZ based analysis**

The laptop computer should cool by itself **(IFR)**.

We want to provide ventilating space inside the laptop so that the heat of the box can be reduced. On the other hand, we don't want to provide ventilating space inside the laptop, as that would increase the size of the laptop **(Contradiction)**.

We want to use multiple cooling fans inside the laptop. Using multiple cooling fans inside the laptop will increase the size of the laptops. But we don't want to increase the size of the laptop **(Contradiction)**.

The invention uses air-ventilating holes through the keyboard **(Principle-31: Hole)**.

**2.6 Keyboard having an integral heat pipe (Patent 6215657)**

**Background problem**

The processor of the computer generates a lot of heat during operation. It's necessary to transfer the heat from the IC for smooth operation of the computer especially for portable computers, which does not provide enough ventilation for heat elimination.

There are various alternative solutions used to solve this problem. (i) Use a metal plate as the heat transfer media, (ii) use the keyboard support plate as the heat transfer media, (iii) heat sinks having integral fans etc. Each having some advantages and disadvantages.

**Solution provided by the invention**

Rakesh Bhatia invented a keyboard having an integral heat pipe (Patent 6215657, assigned to Intel corporation, Apr 01) for heat transmission. The heat pipe provides structural support to the keyboard and is thermally coupled to one or more heat generating components within the computer system. The heat pipe replaces the prior art keyboard support plates. The heat pipe comprises several heat pipes which are arranged in a parallel configuration. Each heat pipe is divided by sidewalls and is sealed to contain a two-phase vaporizable liquid, which serves as the working fluid for the heat pipe.

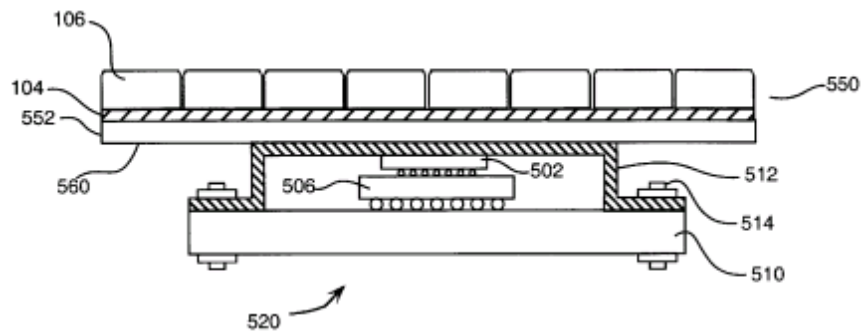

**TRIZ based solution**

The invention uses heat pipes to transfer heat from inside **(Principle-31: Hole)**.

**2.7 Robust keyboard for Public use (Patent 6377246)**

**Background problem**

There are a lot of places where public keyboards are provided for telephones, e-mails, data access, internet and other purposes. A keyboard in a public environment is affected by various weather conditions, rough use and vandalism. That is why a public-use keyboard needs to be robust.
One solution is to keep the keyboard in a metal drawer under lock and key. Allow the user after providing valid authentication such as credit card information. This has two drawbacks, first, the cost of the drawer and locking mechanism adds to the value of the keyboard, second, this does not make the keyboard really public. There is a need for developing a low cost public-use keyboard.

**Solution provided by the invention**

Ronald Wild invented a public-use keyboard (Patent 6377246, Assigned to Lucent Technologies, April 02) which is simple, robust and inexpensive. As per the invention, the keyboard has a keyboard membrane having raised regions on its upper side for the keyboard characters. The keyboard membrane can be formed from a layer of flexible resilient, elastometric material. Alternatively, the keyboard can have a wire braid covering the keyboard membrane that functions as a flexible armor to protect the underlying keyboard. Additionally the keyboard can have a water resistant cover to prevent from liquid like rain, drinks etc.

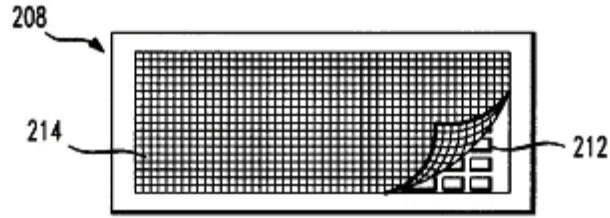

**TRIZ based solution**

The keyboard should protect itself from misuse, attack and vandalism **(Ideal Final Result)**.

The invention uses a thin water resistant cover on the keyboard to protect it from liquids **(Principe-30: Thin and flexible)**.

The inventive method covers the keyboard with wired membrane, which gives protection from physical force **(Principle-30: Thin and flexible)**.

**2.8 Computer keyboard enhancement kit (Patent 6382854)**

**Background problem**

The dispersed structure of the keys on a keyboard makes it difficult for a beginner or a school child to find particular characters on the keyboard. Besides, only capital letter labels confuses a child to find the small letters in the keyboard. There is a need to enhance the keyboard for making the keys easily searchable by small children.

**Solution provided by the invention**

Morelos disclosed a method of detachable key replacements (Patent 6382854, May 2002). The key replacements are attached to the top surface of a transparent plastic jacket that is positioned on top of their corresponding letter keys on the computer keyboard.

The detachable keys will have more shapes and colors than the letter keys of a standard keyboard to make it easy for small children. The shapes and colors follow a specific pattern scheme wherein the letters of the same hand and row keys have key replacements with the same color but with varied shapes.

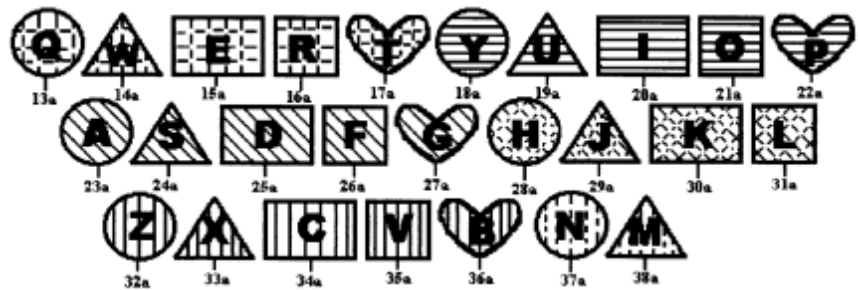

**TRIZ based analysis**

The invention makes use of colors to make the keys prominent **(Principle-32: Color Change)**.

Use replaceable key tops instead of building special keyboards for children **(Principle-34: Discard and Recover)**.

**2.9 Keyboard with detachably attached unit having multimedia key function (Patent 6545668)**

**Background problem**
The multimedia keyboards are typically larger as they contain a few special keys for different kinds of multimedia functions. As the keys are primarily built into the keyboard the keyboard remains larger even if the multimedia functionalities are not used. We need a mechanism to make the keyboard smaller when the multimedia functionalities are not used.

**Solution provided by the invention**
Hayama invented a keyboard (patent 6545668, assigned to Fujitsu Takamisawa, April 2003) which contains an auxiliary keyboard for multimedia functions which is detachable from the main keyboard. When the multimedia key functions are not utilized the auxiliary keyboard unit can be detached from the main keyboard unit to save space.

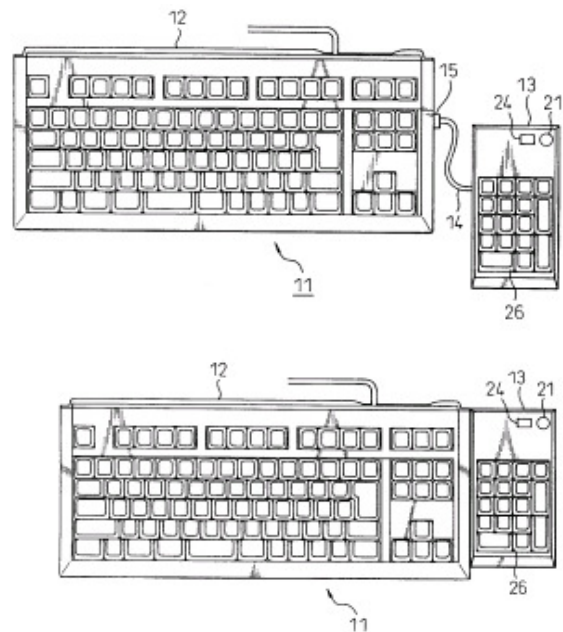

According to the invention the auxiliary keyboard comprises of ten keys for different multimedia functions. The multimedia function definitions are stored in a table in the main computer, which is read by the CPU when powered on.

**TRIZ based analysis**

The same keyboard should provide multimedia functionalities without needing additional space **(desired result)**.

The invention includes multimedia functions in a small keyboard and uses the small keyboard as an attachable unit to the main keyboard **(Principle-15: Dynamize)**.

**2.10 Keyboard with draining unit (Patent 6610944)**

**Background problem**

Most of us have experienced events like coffee or other liquid falling on the keyboard. The conventional structure of keyboard is very inconvenient to clean especially when some liquid like coffee or cold drinks have fallen on it. Once the liquid flows into the keyboard, it spreads into the main body of the keyboard and damages various circuits.

**Solution provided by the invention**

Lee et al. invented a keyboard with a draining unit (US Patent 6610944, Assigned to LG Electronics, Aug 2003) which is capable of draining liquid or a foreign material introduced into a keyboard. The draining unit and the draining structure is simple and does not use any sophisticated or expensive material. The invention discloses three draining systems. The first draining system drains out the keyboard assembly. The second and third draining systems are coupled with the bottom plate. This draining system drains any liquid that falls on the keyboard thereby preventing various circuit parts from being damaged.

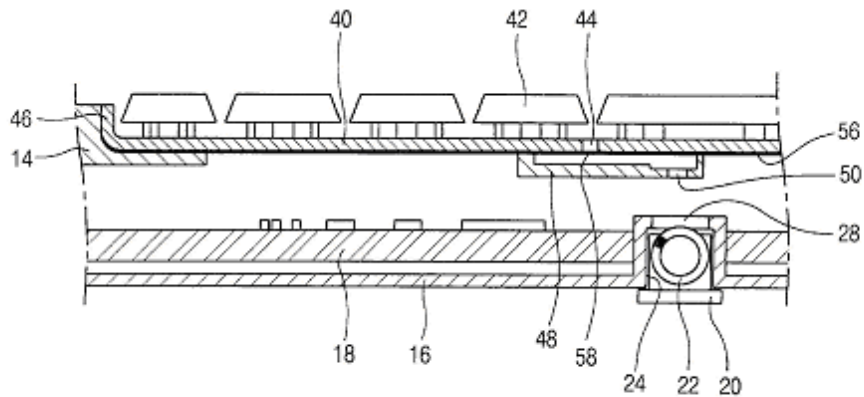

**TRIZ based analysis**

There are several ways to remove liquid from a keyboard. One solution is to use waterproof materials **(Principle-8: Counterweight)**, which are very expensive. Another is to keep absorbents inside the keyboard **(Principle-9: Prior Counteraction)**, which is not very suitable in this case.

This invention uses a simple solution by making a hole in the keyboard for the liquid to drain out **(Principle-31: Hole)**.

## Reference:

1. US Patent 5600313, "Computer keyboard", invented by Lorri Freedman, issued Feb 1997

2. 5767594, Power saving method for computer keyboard. The keyboard goes to sleep mode when not used for a predefined time., Andrew Cheng, assigned to Holtek Microelectronics, June 98

3. 5864334, Computer keyboard with switchable typing/cursor control modes, Sellers, Assigned to Compaq, Jan 99

4. 5978215, Slidable compartments above the keyboard for enhanced cooling mechanism, Chiu et al., assigned to IBM, Nov 99

5. 5982615, The ventilating holes on a keyboard for better cooling system., Kwang-Ho Song, assigned to Samsung Electronics, Nov 99

6. 6215657, Keyboard having a heat pipe for heat transmission., Rakesh Bhatia, assigned to Intel corporation, Apr 01

7. 6377246, Robust keyboard for public use, Ronald Wild, Assigned to Lucent Technologies, April 02

8. 6382854, Detachable key tops for easy recognition of children., Morelos, May 2002

9. 6545232, Light permeable keyboard, Huo-Lu, Sunrex Technology, April 2003

10. 6610944, Keyboard having a draining unit for the liquid to drain out., Lee et al., Assigned to LG Electronics, Aug 2003

11. US Patent and Trademark Office (USPTO) site, http://www.uspto.gov/